\documentclass[12pt,a4paper,english]{article}

\usepackage[utf8]{inputenc}
\usepackage[T1]{fontenc}
\usepackage[english]{babel}

\usepackage[usenames,dvipsnames]{xcolor}
\usepackage{pgf,tikz}
\usetikzlibrary{arrows}
\usetikzlibrary{decorations.markings,decorations.pathmorphing,patterns}

\usepackage[format=hang,font=small]{caption}
\usepackage{booktabs,bm,multirow,subfig}%
\usepackage{enumitem}
\usepackage[affil-it,auth-sc]{authblk}

\usepackage{amsmath,amsfonts,amssymb,amsthm}%
\DeclareMathAlphabet{\mathcalligra}{OT1}{pzc}{m}{it}

\usepackage
{graphicx}

\graphicspath{{Images/}}

\renewcommand{\phi}{\varphi}
\renewcommand{\theta}{\vartheta}
\renewcommand{\epsilon}{\varepsilon}

\newcommand{\field}[1]{\mathbb{#1}} 
\newcommand{\R}{\field{R}}

\newcommand{\T}{\mathcal{T}}
\newcommand{\CC}{\mathcal{C}}
\newcommand{\Clip}{\mathcal{C}^\mathrm{Lip}}
\newcommand{\EE}{\mathcal{E}}
\newcommand{\DD}{\mathcal{D}}

\newcommand{\Dd}{\mathcalligra{d}}
\newcommand{\Ds}{\mathcal{D}_\mathrm{sh}}
\newcommand{\vm}{v_\mathrm{min}}

\DeclareMathOperator{\midd}{middle}
\DeclareMathOperator{\conv}{conv}
\DeclareMathOperator{\inter}{int}
\newcommand{\pvector}[1]{\begin{pmatrix}#1\end{pmatrix}}
\newcommand{\edge}[2]{\overline{#1 #2}}

\newcommand{\scal}[2]{\left\langle #1,#2\right\rangle}
 
\newcommand{\dert}[1]{\dot{#1}} 

\newcommand*{\diff}{\mathop{}\!\mathrm{d}}
\newcommand{\dd}{\diff}

\newtheorem{theorem}{Theorem}

\theoremstyle{definition}

\numberwithin{equation}{section}

\title{Stasis domains and slip surfaces in the locomotion of a bio-inspired two-segment crawler}

\author[1]{Paolo Gidoni}
\author[1,2,\footnote{Corresponding author -- e-mail address:\,\texttt{desimone@sissa.it} -- phone:\,+39\,040\,3787\,455}]{Antonio DeSimone}

\affil[1]{\small{SISSA -- International School for Advanced Studies, via Bonomea 265, 34136 Trieste, Italy.} \smallskip}
\affil[2]{\small{GSSI -- Gran Sasso Science Institute, viale Francesco Crispi 7, 67100 L'Aquila, Italy.} \smallskip}

\date{}

\begin{document}
\maketitle

\begin{abstract}
We formulate and solve the locomotion problem for a bio-inspired crawler consisting of two active elastic segments (i.e., capable of changing their rest lengths), resting on three supports providing directional frictional interactions.
The problem consists in finding the motion produced by a given, slow actuation history.

By focusing on the tensions in the elastic segments, we show that the evolution laws for the system are entirely analogous to the flow rules of elasto-plasticity.
In particular,  sliding of the supports and hence motion cannot occur when the tensions are in the interior of certain convex regions (stasis domains), while support sliding (and hence motion) can only take place when the tensions are on the boundary of such regions (slip surfaces).

We solve the locomotion problem explicitly in a few interesting examples. In particular, we show that, for a suitable  range of the friction parameters, specific choices of the actuation strategy can lead to net displacements also in the direction of higher friction.

\end{abstract}


\section{Introduction}

Research on biological and bio-inspired locomotion, aimed at understanding and replicating motor abilities of animals capable of propelling effectively in environments where standard locomotion
strategies fail (e.g., those based on wheels), is receiving increasing attention, starting from the seminal work by Hirose \cite{Hir93}. 

A promising area where interesting applications are envisaged  is medical endoscopy, through the development of miniaturised biomedical robotic tools \cite{Iku03, Men03}. Here, the need for non-standard
bio-inspired solutions comes from size constraints (which make devices based on engine-powered shafts and cogwheels unfeasible) and from the challenge of extracting propulsive forces 
from the frictional interactions with soft biological tissues in a non-invasive way. Drawing inspiration from the locomotion strategies of worms (e.g., {\it Lumbricus terrestris}), and from the
anchoring abilities of parasites and larvae, artificial bio-mimetic crawlers have been conceived, manufactured, and analysed.
The system considered in \cite{Men06}, a prototypical example, consists of several elastic segments that can actively change their rest lengths (thanks to shape-memory-alloy wires, actuated with electric currents via Joule heating), and supported by hook-shaped elements that give a directional character to the frictional interactions with the environment (as in ``hairy’’ surfaces, characterised by low friction when sliding occurs ``along the grain’’, and by high friction when sliding occurs in the opposite direction, ``against the grain’’). Similar systems have been investigated in the context of the more general robotics literature \cite{BoletA15,BorFigChe14,SteBeh12,ZimetA09}, or in models for the propulsion of crawling cells \cite{RecTru15}.
In spite of the many interesting results contained in these studies, several open issues remain, even at the level of theoretical analysis. In particular, a detailed understanding of the general relation between actuation history, elastic tensions developed in the segments, and observable motions is still missing. In \cite{Men06}, for example, the simplifying assumption is made that motion in the high friction direction is forbidden, i.e., no back-sliding occurs. As a consequence, motion can only take place in one direction, the one of low resistance.

Inspired by these developments, and building upon previous work by our group \cite{DeSTat12,DeSetA13,DeSGidNos15,GidNosDeS14,NosDeSTat13,NosDeS14}, we consider in this paper a model crawler lying on a horizontal surface, consisting of two linearly elastic segments that can actively change their rest length, and subject to horizontal frictional forces at their ends mimicking a directional frictional contact. 
We formulate the locomotion problem for this system in the regime of slow (quasi-static) actuation,  in which inertial forces can be neglected, an actuation history is prescribed by assigning the time evolution  of the rest length of the segments, and we solve for the resulting motion. In this way, we extend the results obtained in \cite{GidNosDeS14}, where the behaviour of a one-segment crawler was analysed. The increased complexity of the two segment crawler requires a methodological change. Indeed, we solve the problem by showing that the behaviour of the system is governed by the tensions arising in the elastic segments, and that the resulting laws of motion are entirely analogous to the flow rules typical of elasto-plasticity.
In particular, there are convex domains in the plane of the internal tensions (stasis domains, the analog of elastic domains) corresponding to which no sliding of the supports can take place.
Only when the tensions reach the boundaries of these domains (slip surfaces, the analog of yield surfaces), sliding of the supports, and hence  motion of the segments can occur.

We solve the locomotion problem in a few interesting examples. In particular, we show that, for a suitable  range of the friction parameters, specific choices of the actuation strategy can lead to net displacements also in the direction of higher friction.
This last remark shows that, provided that the system is complex enough (i.e., it is made of at least two independent segments), it is  not only motile (i.e., it can exhibit non-zero net displacements in at least one direction), but it is in fact controllable (it can move in both directions).  It would be interesting to investigate whether a similar controllable motility scenario can also emerge in a different but related context, namely, forced brownian particles in a non-symmetric potential (forced thermal ratchets, see \cite{Magnasco,Recho}) which have been used as a model for the motion of motor proteins along microtubules and actin filaments.

The rest of the paper is organised as follows. In Section \ref{sec:mech} we present our model of crawler and formulate the motility problem, introducing a necessary dimensional reduction that we discuss in detail in Section \ref{sec:shape}. In Section \ref{sec:law} we study the associated stasis domains and deduce the laws of motion, which are discussed in Section \ref{sec:disc}. Here, to better illustrate the situation, we construct and analyse two periodic motility strategies, generating displacement in opposite directions.

\section{The crawler: formulation of the problem} \label{sec:mech}

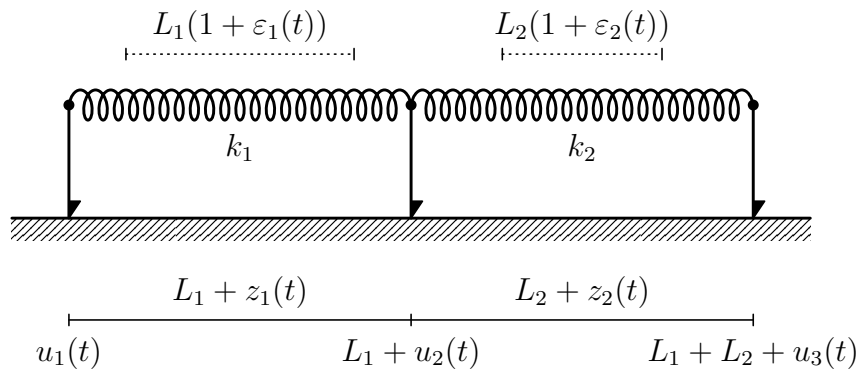
\begin{figure}
\centering
\begin{tikzpicture}[line cap=round,line join=round,>=triangle 45,x=1.0cm,y=1.0cm, line width=1.1pt,scale=1.5]
\clip(0.,-1.5) rectangle (8.,2.);

\draw[line width=0.6pt] (1,-0.9)-- (7,-0.9);
\draw[line width=0.6pt] (1,-0.85)-- (1.0,-0.95) node[anchor=north] {$u_1(t)$};
\draw[line width=0.6pt] (4,-0.85)-- (4.0,-0.95) node[anchor=north] {$L_1+u_2(t)$};
\draw[line width=0.6pt] (7,-0.85)-- (7.0,-0.95) node[anchor=north] {$L_1+L_2+u_3(t)$};
\draw (2.5,-0.9) node[anchor=south] {$L_1+z_1(t)$};
\draw (5.5,-0.9) node[anchor=south] {$L_2+z_2(t)$};
\draw[dotted, line width=0.6pt] (1.5,1.45)-- (3.5,1.45);
\draw[line width=0.6pt] (1.5,1.4)-- (1.5,1.5);
\draw[line width=0.6pt] (3.5,1.4)-- (3.5,1.5);
\draw[dotted, line width=0.6pt] (4.8,1.45)-- (6.2,1.45);
\draw[line width=0.6pt] (4.8,1.4)-- (4.8,1.5);
\draw[line width=0.6pt] (6.2,1.4)-- (6.2,1.5);
\draw (2.5,1.45) node[anchor=south] {$L_1(1 +\epsilon_1(t))$};
\draw (5.5,1.45) node[anchor=south] {$L_2(1+ \epsilon_2(t))$};
\draw (2.5,0.85) node[anchor=north] {$k_1$};
\draw (5.5,0.85) node[anchor=north] {$k_2$};
\draw (1.,1.)-- (1.,0.);
\draw[decoration={aspect=0.5, segment length=2mm, amplitude=2mm,coil},decorate] (1,1.)-- (4.,1.);
\draw (4.,1.)-- (4.,0.);
\draw [decoration={aspect=0.5, segment length=2mm, amplitude=2mm,coil},decorate](4,1.)-- (7.,1.);
\draw (7.,1.)-- (7.,0.);
\draw (0.5,0.)-- (7.5,0.);
\draw [fill=black] (1.,1.) circle (1pt);
\draw [fill=black,line width=0.6pt] (1.,0.) --  (1,0.15) -- (1.1,0.15) --(1,0);
\draw [fill=black,line width=0.6pt] (4.,0.) --  (4,0.15) -- (4.1,0.15) --(4,0);
\draw [fill=black,line width=0.6pt] (7.,0.) --  (7,0.15) -- (7.1,0.15) --(7,0);
\draw [fill=black] (7.,1.) circle (1pt);
\draw [fill=black] (4.,1.) circle (1pt);
\fill [pattern = north east lines] (0.5,0) rectangle (7.5,-0.2);
\end{tikzpicture}
\caption{The model of our crawler. The dotted lines represent the rest lengths of the two springs.}
\label{fig:crawler}
\end{figure}

We are interested in the motion of one-dimensional crawlers such as that represented in Figure \ref{fig:crawler}. The crawler is composed of two adjacent rods, identified in the reference configuration by the segments $[X_1,X_2]$ and $[X_2,X_3]$. We assume $X_1=0$, $X_2=L_1$ and $X_2=L_1+L_2$, so that $L_1$ and $L_2$ are the reference lengths of the two rods. A point $X$ of the crawler is mapped to the point $x=\chi(X,t)$ in the deformed configuration and thus its displacement is $u(X,t)=\chi(X,t)-X$. It is useful to set $u_1(t)=u(X_1,t)$, $u_2(t)=u(X_2,t)$ and $u_3(t)=u(X_3,t)$.

We denote the derivatives with respect to space and time with a prime and a dot, respectively, 
\begin{align}
u'(X,t)=\frac{\partial }{\partial X} u(X,t) && \dert u(X,t)=\frac{\partial }{\partial t} u(X,t)
\end{align}

The crawler interacts with the substrate only through three rigid legs located at $X_1$, $X_2$ and $X_3$. These interactions are described by the (directional) friction law
\begin{equation}
F_i(t)=F(X_i,t)\in 
\begin{cases}
\{F_-\} &\text{if $\dert u_i(t)<0$}\\
[-F_+,F_-] &\text{if $\dert u_i(t)=0$}\\
\{-F_+\} &\text{if $\dert u_i(t)>0$}\\
\end{cases}
\end{equation}
where $i=1,2,3$. We assume that 
\begin{equation}
F_->F_+>0
\label{cond:Fnot1}
\end{equation}
This means that the absolute value of the friction force is not constant and depends on the direction of motion; moreover the coordinates are chosen so that negative velocities generate  greater friction.

The two rods are assumed to be elastic, with stiffnesses $k_1$,$k_2$, and subject to an active distortion $\epsilon_0(X,t)$. We assume that the distortion is uniform along each rod so that
\begin{equation}
\epsilon_0(X,t)=\begin{cases}
 \epsilon_1(t) &\text{if $X\in (0,L_1)$}\\
\epsilon_2(t) &\text{if $X\in (L_1,L_1+L_2)$}
\end{cases}
\end{equation}
The rest length of the two rods is thus $(1+\epsilon_1(t))L_1$ and $(1+\epsilon_2(t))L_2$, respectively.

\subsection{Internal energy and dissipation}

For our analysis it is useful to describe the state of the crawler with two parameters $z=(z_1, z_2)^t$ associated with its shape and a parameter $y$ that identifies its position.
More precisely, we set
\begin{align}
z_1(t)=u_2(t)-u_1(t) &&  z_2(t)=u_3(t)-u_2(t) && y(t)=u_2(t)
\end{align}

As we will show, the internal energy of the crawler depends only on its shape and the dissipation, in almost all circumstances, can be expressed as a function of just the shape change $\dert z$ through some minimality considerations. This will allow us to model the crawler as a rate independent dissipative system with quadratic (positive definite) energy, a situation well studied in elastoplasticity \cite{Mie04, Mie05}.

The stored energy of the crawler is given by
\begin{align}
\EE&=\frac{k_1}{2}\int_0^{L_1}(u'(X,t)-\epsilon_1)^2\dd X +\frac{k_2}{2}\int_{L_1}^{L_1+L_2}(u'(X,t)-\epsilon_2)^2\dd X \notag \\
&=\frac{k_1 L_1}{2}\left[\frac{u_2(t)-u_1(t)}{L_1}-\epsilon_1(t)\right]^2+
\frac{k_2 L_2}{2}\left[\frac{u_3(t)-u_2(t)}{L_2}-\epsilon_2(t)\right]^2\notag\\
&=\frac{1}{2}\scal{Az(t)}{z(t)}-\scal{\ell(t)}{z(t)} +c(t)
\end{align}
where we have used the fact that minimal energy leads to $X\mapsto u'(x,t)$ constant along each of the two rods, and we have set
\begin{gather*}
A=\begin{pmatrix}
\frac{k_1}{L_1}& 0\\
0 &  \frac{k_2}{L_2}
\end{pmatrix} \qquad\qquad
\ell(t)=\begin{pmatrix}
k_1 \epsilon_1(t)\\
k_2 \epsilon_2(t)
\end{pmatrix}\\
c(t)=\frac{k_1 L_1 \epsilon_1(t)^2}{2}+\frac{k_2 L_2 \epsilon_2(t)^2}{2}
\end{gather*}
We thus see that, for a prescribed active distortion $\epsilon (t)$, the internal energy of the crawler depends only on time and on the shape $z(t)$, allowing us to write from now on $\EE=\EE(t,z(t))$.

The dissipation produced by the displacement $u_i\mapsto u_i+v_i$ of a single contact point is
\begin{equation}
\Dd(v_i)= v_i^+F_+ - v_i^-F_- 
\end{equation}
where
\begin{align}
&v_i^+=\begin{cases}
v_i & \text{if $v_i\geq 0$}\\
0 & \text{if $v_i< 0$}\\
\end{cases}
& \text{and}
&&v_i^-=\begin{cases}
v_i & \text{if $v_i\leq 0$}\\
0 & \text{if $v_i> 0$}\\
\end{cases}
\end{align}
and therefore the dissipation produced by a shape change $z\mapsto z+w$ and a position change $y\mapsto y+v$ is
\begin{equation}
\DD(w,v)= \Dd(v-w_1) +\Dd(v) +\Dd(v+w_2)
\end{equation}
We observe that $\DD$ is convex and positively homogeneous of degree $1$.

For any fixed shape change $w=\bar w$, the function $v \mapsto \DD(\bar w, v)$ is convex and coercive, as it is the sum of convex and coercive functions of $v$. We now show that, under the additional hypothesis
\begin{equation}
F_- \neq 2F_+
\label{cond:Fnot2}
\end{equation}
it has an unique minimum value, attained at $v=\vm (\bar w)$.

First of all we observe that $\DD(\bar w, \cdot)$ is differentiable everywhere except on the finite set $ \{\bar w_1, 0, -\bar w_2 \}$. The asymmetry of the friction \eqref{cond:Fnot1} and the additional assumption \eqref{cond:Fnot2} ensure  that 
\begin{equation}
\frac{\partial \DD(\bar w, v)}{\partial v}\neq 0 \qquad\text{for every $\bar w\in \R^2$ and every $v\in \R\setminus \{\bar w_1, 0, -\bar w_2 \} $}
\end{equation}
Hence $\DD(\bar w, \cdot)$ has an unique minimum attained at $v=\vm (\bar w)\in \{\bar w_1, 0, -\bar w_2 \}$.
With simple considerations on the sign of the derivative we can determine the exact value of $\vm$. Precisely
\begin{equation}
\vm (\bar w)=\begin{cases} 
\max \{\bar w_1, 0, -\bar w_2 \} & \text{if $F_- > 2F_+$}\\
\midd (\bar w_1, 0, -\bar w_2) & \text{if $2F_+> F_- >F_+$}
\end{cases}
\end{equation}
where we have introduced a \lq $\midd$\rq\ function that returns
\begin{itemize}
\item if its three arguments have all different values, the one with the middle value; 
\item if at least two arguments have the same value, that value.
\end{itemize}
More pragmatically, we order the triplet $(\bar w_1, 0, -\bar w_2)$ and pick the middle element.

\begin{figure}[tb]
\centering
\subfloat[][\emph{Case $F_- > 2F_+$}.]
{\begin{tikzpicture}[line cap=round,line join=round,x=1.0cm,y=1.0cm,scale=0.3,line width=0.8pt]
\draw[->,color=black] (-9.,0.) -- (9.,0.);
\draw[->,color=black] (0.,-9.) -- (0.,9.);
\clip(-10.,-10.) rectangle (10.,10.);
\draw [line width=1.2pt,color=Green,domain=0.0:8.0] plot(\x,{(-0.-1.*\x)/1.});
\draw [line width=1.2pt,color=Green] (0.,0.) -- (0.,8.);
\draw [line width=1.2pt,color=Green,domain=-8.0:0.0] plot(\x,{(-0.-0.*\x)/-1.});
\draw (-5,4) node{$A_1$};
\draw (4.,3) node{$A_2$};
\draw (-3.,-3.5) node {$A_3$};
\draw (0.,8.5) node[anchor= west] {\footnotesize $w_2$};
\draw (8.5,0) node[anchor=north] {\footnotesize $w_1$};
\draw [dashed,color=red] (-8.,-1.) -- (1.,-1.) -- (1.,8.);
\draw [dashed,color=red] (-8.,-2.) -- (2.,-2.) -- (2.,8.);
\draw [dashed,color=red] (-8.,-3.) -- (3.,-3.) -- (3.,8.);
\draw [dashed,color=red] (-8.,-4.) -- (4.,-4.) -- (4.,8.);
\draw [dashed,color=red] (-8.,-5.) -- (5.,-5.) -- (5.,8.);
\draw [dashed,color=red] (-8.,-6.) -- (6.,-6.) -- (6.,8.);
\draw (-5.,2) node {\footnotesize $\vm(w)=0$};
\draw (3.5,-7.5) node[color=red] {\footnotesize $\vm(w)>0$};
\end{tikzpicture}} \quad
\subfloat[][\emph{Case $2F_+> F_- >F_+$}.]
{\begin{tikzpicture}[line cap=round,line join=round,x=1.0cm,y=1.0cm,scale=0.3,line width=0.8pt]
\draw[->,color=black] (-9.5,0.) -- (9.5,0.);
\draw[->,color=black] (0.,-9.5) -- (0.,9.5);
\clip(-10.,-10.) rectangle (10.,10.);
\draw (5,5.) node {$B_3$};
\draw (-2.,6.) node {$B_2$};
\draw (-6.,2.5) node {$B_1$};
\draw (-5.,-5.) node {$B_6$};
\draw (2.,-6.) node {$B_5$};
\draw (6.,-2.5) node {$B_4$};
\draw [line width=1.2pt,color=Green,domain=-8.:8.] plot(\x,{(-0.-1.*\x)/1.});
\draw [line width=1.2pt,color=Green] (0.,-8.) -- (0.,8.);
\draw [line width=1.2pt,color=Green,domain=-8.:8.] plot(\x,{(-0.-0.*\x)/1.});
\draw (0.,8.5) node[anchor= west] {\footnotesize $w_2$};
\draw (8.5,0) node[anchor=north] {\footnotesize $w_1$};;
\draw [dashed,color=blue,domain=-8.0:-1.0] plot(\x,{(-6.-0.*\x)/-6.});
\draw [dashed,color=blue,domain=-8.0:-2.0] plot(\x,{(-10.-0.*\x)/-5.});
\draw [dashed,color=blue,domain=-8.0:-3.0] plot(\x,{(-12.-0.*\x)/-4.});
\draw [dashed,color=blue,domain=-8.0:-4.0] plot(\x,{(-12.-0.*\x)/-3.});
\draw [dashed,color=blue,domain=-8.0:-5.0] plot(\x,{(-10.-0.*\x)/-2.});
\draw [dashed,color=blue,domain=-8.0:-6.0] plot(\x,{(-6.-0.*\x)/-1.});
\draw [dashed,color=blue] (-6.,6.) -- (-6.,8.);
\draw [dashed,color=blue] (-5.,5.) -- (-5.,8.);
\draw [dashed,color=blue] (-4.,4.) -- (-4.,8.);
\draw [dashed,color=blue] (-3.,3.) -- (-3.,8.);
\draw [dashed,color=blue] (-2.,2.) -- (-2.,8.);
\draw [dashed,color=blue] (-1.,1.) -- (-1.,8.);
\draw [dashed,color=red,domain=1.0:8.0] plot(\x,{(-6.-0.*\x)/6.});
\draw [dashed,color=red,domain=2.0:8.0] plot(\x,{(-10.-0.*\x)/5.});
\draw [dashed,color=red,domain=3.0:8.0] plot(\x,{(-12.-0.*\x)/4.});
\draw [dashed,color=red,domain=4.0:8.0] plot(\x,{(-12.-0.*\x)/3.});
\draw [dashed,color=red,domain=5.0:8.0] plot(\x,{(-10.-0.*\x)/2.});
\draw [dashed,color=red,domain=6.0:8.0] plot(\x,{(-6.-0.*\x)/1.});
\draw [dashed,color=red] (1.,-1.) -- (1.,-8.);
\draw [dashed,color=red] (2.,-2.) -- (2.,-8.);
\draw [dashed,color=red] (3.,-3.) -- (3.,-8.);
\draw [dashed,color=red] (4.,-4.) -- (4.,-8.);
\draw [dashed,color=red] (5.,-5.) -- (5.,-8.);
\draw [dashed,color=red] (6.,-6.) -- (6.,-8.);
\draw (-5.,-7) node {\footnotesize $\vm(w)=0$};
\draw (5.,3) node {\footnotesize $\vm(w)=0$};
\draw (6,-9) node[color=red] {\footnotesize $\vm(w)>0$};
\draw (-6,9) node[color=blue] {\footnotesize $\vm(w)<0$};
\end{tikzpicture}} 
\caption{Contour plot (dashed) of the function $\vm(w)$ for different choices of the friction parameters.}
\label{fig:vm}
\end{figure}
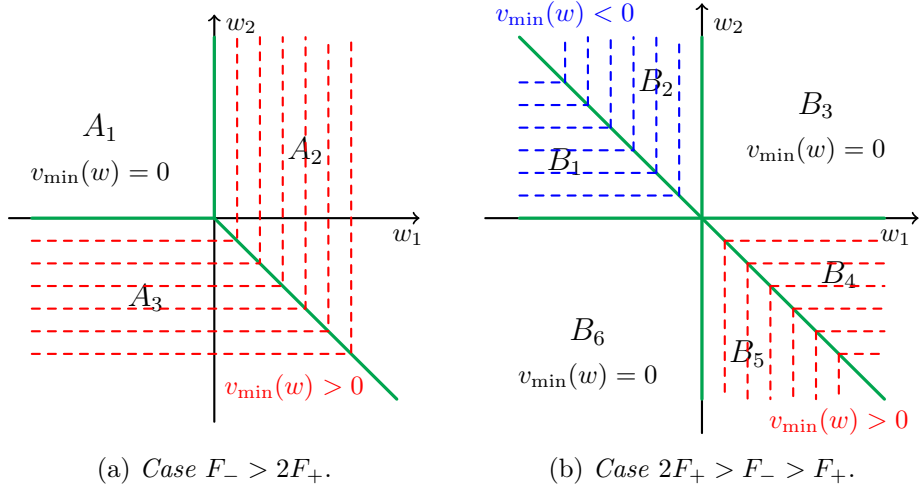

 We observe that the $\vm$ is positively homogeneous of degree 1; its behaviour according to the values of the friction force is illustrated in Figure~\ref{fig:vm}.

\subsection{The variational inequality}

We assume that the actuation history is slow enough (quasi-static) that inertial forces can be neglected. The evolution of the system is thus governed by the balance of forces, namely, by the fact that the sum of frictional resistance forces $F_i$ and elastic restoring forces is zero. This is expressed in abstract form by \eqref{eq:SF} below and, more concretely, by the following three equations 
\begin{equation} \label{eq:forcebalance}
\begin{cases}
F_1+\frac{k_1}{L_1}(z_1-L_1\epsilon_1)=0  \\
F_2-\frac{k_1}{L_1}(z_1-L_1\epsilon_1)+\frac{k_2}{L_2}(z_2-L_2\epsilon_2)=0 \\
F_3-\frac{k_2}{L_2}(z_2-L_2\epsilon_2)=0 \\
\end{cases}
\end{equation}
An alternative way to write system \eqref{eq:forcebalance} is to cast it in the form of a variational inequality (see \cite{Mie04,Mie15}). Doing this will enable us to exploit some known results on the evolution of rate independent systems. Therefore, we will write the laws governing the evolution of our system as a variational inequality first, and show later that this formulation leads to \eqref{eq:SF} and \eqref{eq:forcebalance}.

For a given external load $\ell (t)$, the evolution $z(t),y(t)$ of our system is obtained as a solution of the variational inequality
\begin{equation}
\scal{Az(t)-\ell (t)}{w-\dert z(t)}+\DD(w,v)-\DD(\dert z(t),\dert y(t))\geq 0 \tag{VI}
\label{eq:VI}
\end{equation}
for every $(w,v)\in \R^2\times \R$.
In particular this must hold for $w=\dert z(t)$, for which we get
\begin{equation}
\DD(\dert z(t),v)-\DD(\dert z(t),\dert y(t))\geq 0 \quad \text{for every $v\in\R$}
\end{equation}
This is equivalent to set
\begin{equation}
\dert y(t)=\vm (\dert z(t))
\label{eq:yz_coupling}
\end{equation}
We can use this fact to reduce the dimension of the problem associated to the variational inequality \eqref{eq:VI}, leading to
\begin{equation}
\scal{Az(t)-\ell (t)}{w-\dert z(t)}+\Ds(w)-\Ds(\dert z(t))\geq 0 \quad \text{for every $w\in \R^2$} \tag{RVI}
\label{eq:RVI}
\end{equation}
where $\Ds$ is the \lq\lq shape-restricted\rq\rq\ dissipation, i.e.~the dissipation after minimization with respect to translations of the crawler,
\begin{equation}
\Ds(w)=\DD(w,\vm(w))
\end{equation}
This allows us to study the system for the shape changes alone and then recover the displacement $y(t)$ of the crawler through the relationship \eqref{eq:yz_coupling}.

Before discussing existence and uniqueness of the solutions for our problem, let us notice that $\Ds$ is convex (and therefore continuous) and positively homogeneous of degree $1$.
To show this, we recall that $w\mapsto \vm(w)$ is positively homogeneous of degree $1$. Hence, for $\lambda>0$
\begin{equation*}
\Ds(\lambda w)=\DD(\lambda w, \vm(\lambda w))=\DD(\lambda w, \lambda \vm(w))
=\lambda\DD( w, \vm(w))=\lambda \Ds(w)
\end{equation*}
Regarding the convexity of $\Ds$, we observe that for every  $0\leq \lambda \leq 1$, writing $w_\lambda=\lambda w + (1-\lambda \bar w)$, we have
\begin{align}
\lambda \Ds(w)+(1-\lambda)\Ds (\bar w) &\geq 
\lambda \DD(w,\vm(w))+(1-\lambda)\DD (\bar w,\vm(\bar w))\notag\\
&\geq \DD(w_\lambda, \lambda\vm(w)+(1-\lambda)\vm(\bar w))\notag\\
&\geq \DD(w_\lambda, \vm(w_\lambda))=\Ds(w_\lambda)
\end{align}

Let us recall that 
the subdifferential of $\Ds$ in $\bar w$ is defined as
\begin{equation*}
\partial \Ds(\bar w)=\{\xi\in \R^2 \colon \Ds(w)\geq \Ds(\bar w)+\scal{\xi}{w-\bar w} \text{for every $w\in \R^2$}\}
\end{equation*}
We remark that, strictly speaking, the subdifferential consists of elements of the dual space $(\R^2)^*$, but since we are working with finite dimensional spaces we implicitly adopt the usual identification of the elements of the dual with vectors of the space. 
Setting $C^*=\partial \Ds(0)$, we observe that $C^*$ is convex and satisfies
\begin{equation}
\Ds(w)=\max_{\xi\in C^*} \scal{\xi}{w}
\end{equation}

We have the following result (cf.~\cite[Theorem 2.1]{Mie05}).
\begin{theorem}\label{th:mielke}
Given $\ell \in \CC^1([0,\T],\R^2)$ and $z_0\in A^{-1}(\ell (0)-C^*)$, there exists a unique function $z\in\Clip([0,\T],\R^2)$, with $z(0)=z_0$ and such that the shape-restricted variational inequality \eqref{eq:RVI} is satisfied for almost all $t\in [0,\T]$.
\end{theorem}
We remark that the dimensional reduction that allowed us to pass from $\DD $ to $\Ds$ is necessary to attain uniqueness, since the energy $\EE(t,\cdot)$ is not uniformly convex on $\R^3$, but becomes so if restricted to the shape coordinates $z$. 
When assumption \eqref{cond:Fnot2} does not hold, it is possible to find multiple solutions for problem \eqref{eq:VI}, as shown by the following example.

Let us set $F_-=2F_+$ and assume that, at the initial time $t=0$, the state of the crawler is such that both the springs are in the state of maximum compression, namely
\begin{align*}
\frac{k_1}{L_1}(z_1-L_1\epsilon_1)= -F_- &&
\frac{k_2}{L_2}(z_2-L_2\epsilon_2)= -F_+
\end{align*}
We consider an external load such that, for $t\in[0,\T]$, we have $\dert \epsilon_1(t)>0$ and $\dert \epsilon_2(t)=0$. Under this conditions, the system has infinite solutions, identified by the parameter $\mu\in[0,1]$ and defined by
\begin{align*}
\dert u_1(t)=-\mu L_1\dert \epsilon_1(t) && \dert u_2(t)=\dert u_3(t)=(1-\mu)L_1\dert \epsilon_1(t)
\end{align*}

We also observe that, using the definition of subdifferential, the variational inequality \eqref{eq:RVI} can be restated as 
\begin{equation} \label{eq:SF}
0\in \partial \Ds(\dert z(t))+D_z\EE(t,z(t)) \tag{SF}
\end{equation}
that is called the \emph{subdifferential formulation} of the problem.

As remarked above, since inertial forces can be neglected in the regime of quasi-static actuation, \eqref{eq:SF} is a force balance stating that the sum of dissipative frictional forces $\partial \Ds(\dert z(t))$ and elastic restoring forces $D_z\EE(t,z(t))$ must vanish at all times.



\section{The shape-dependent dissipation} \label{sec:shape}

Our next step is therefore to study the restricted dissipation $\Ds$ and express more explicitly its differential. We consider separately the two cases $F_->2F_+$ and $2F_+>F_->F_+$, since a different behaviour is observed.

\subsection{Case $F_->2F_+$}
\begin{figure}[th]\centering
\begin{tikzpicture}[line cap=round,line join=round, x=1.0cm,y=1.0cm, scale=0.5]
\draw[->,color=black] (-7.,0.) -- (7.5,0.);
\draw[->,color=black] (0.,-7.) -- (0.,7.5);
\clip(-9.,-9.) rectangle (9.,9.);
\fill[line width=1.2pt,color=red,fill=red,fill opacity=0.5] (-1.,1.) -- (2.,1.) -- (-1.,-2.) -- cycle;
\draw [dashed] (0.,1.)-- (-1.,0.);
\draw [dashed] (0.,2.)-- (-2.,0.);
\draw [dashed] (0.,3.)-- (-3.,0.);
\draw [dashed] (0.,4.)-- (-4.,0.);
\draw [dashed] (0.,5.)-- (-5.,0.);
\draw [dashed] (0.,6.)-- (-6.,0.);
\draw [dashed] (0.,1.)-- (1.,-1.);
\draw [dashed] (0.,2.)-- (2.,-2.);
\draw [dashed] (0.,3.)-- (3.,-3.);
\draw [dashed] (0.,4.)-- (4.,-4.);
\draw [dashed] (0.,5.)-- (5.,-5.);
\draw [dashed] (0.,6.)-- (6.,-6.);
\draw [dashed] (1.,-1.)-- (-1.,0.);
\draw [dashed] (-2.,0.)-- (2.,-2.);
\draw [dashed] (-3.,0.)-- (3.,-3.);
\draw [dashed] (-4.,0.)-- (4.,-4.);
\draw [dashed] (-5.,0.)-- (5.,-5.);
\draw [dashed] (-6.,0.)-- (6.,-6.);
\draw [line width=1.2pt,color=Green] (0.,0.) -- (0.,7.);
\draw [line width=1.2pt,color=Green] (0.,0.) -- (-7.,0.);
\draw [line width=1.2pt,color=Green] (0.,0.) -- (7.,-7.);
\draw [line width=1.2pt,color=red] (-1.,1.)-- (2.,1.);
\draw [line width=1.2pt,color=red] (2.,1.)-- (-1.,-2.);
\draw [line width=1.2pt,color=red] (-1.,-2.)-- (-1.,1.);
\draw (-5,4) node{$A_1$};
\draw (4.,3) node{$A_2$};
\draw (-3.,-4.) node {$A_3$};
\draw (2.,-1.) node[color=red, fill=white] {$C^*_A$};
\draw (7,0) node[anchor=north] {$w_1$};
\draw (0.,7.) node[anchor= west] {$w_2$};
\end{tikzpicture}
\caption{Case $F_->2F_+$. The three regions $A_1$, $A_2$ and $A_3$, the contour lines of $\Ds$ (dashed) and its subdifferential at the origin $C^*_A$ (red).}
\label{fig:regA}
\end{figure}
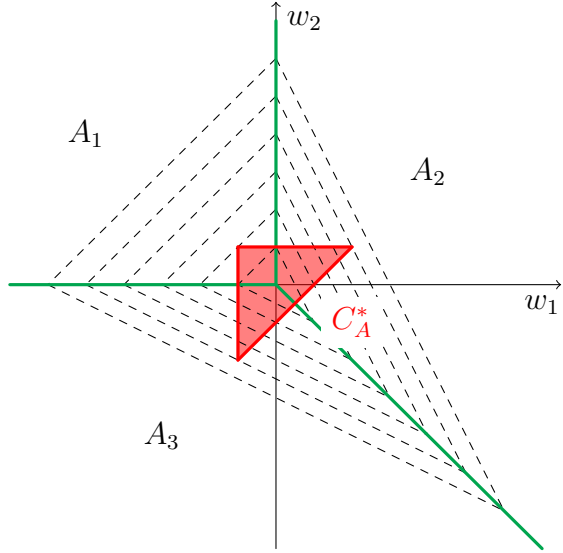

We divide the plane into three regions $A_1$, $A_2$ and $A_3$, as shown in Figure \ref{fig:regA}. 

\begin{enumerate}[label=\textup{($A_{\arabic*}$)}]
\item  This is the region defined by  $w_1 \leq 0\leq w_2$, that implies $\vm(w)=0$ and
\begin{equation*}
\Ds(w)=(-w_1+w_2)F_+=\scal{\alpha_1}{w} \qquad \text{where $\alpha_1=\begin{pmatrix}
-F_+ \\ F_+
\end{pmatrix}$}
\end{equation*}
\item Here we have $w_1\geq 0$ and $-w_2\leq w_1$, so $\vm(w)=w_1$ and
\begin{equation*}
\Ds(w)=(2w_1+w_2)F_+=\scal{\alpha_2}{w} \qquad \text{where $\alpha_2=\begin{pmatrix}
2F_+ \\ F_+
\end{pmatrix}$}
\end{equation*}
\item  Here we have $w_2\leq 0$ and $-w_2\geq w_1$, so $\vm(w)=-w_2$ and
\begin{equation*}
\Ds(w)=(-w_1-2w_2)F_+=\scal{\alpha_3}{w} \qquad \text{where $\alpha_3=\begin{pmatrix}
-F_+ \\ -2F_+
\end{pmatrix}$}
\end{equation*}
\end{enumerate}

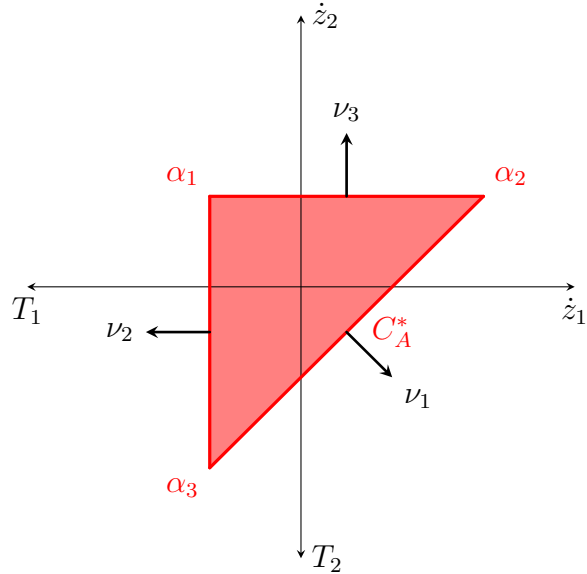
\begin{figure}[tb]\centering
\begin{tikzpicture}[line cap=round,line join=round,x=1.0cm,y=1.0cm,>=stealth,scale=1.2]
\clip(-4.,-4.) rectangle (4.,4.);
\fill[line width=1 pt,color=red,fill=red,fill opacity=0.5] (-1.,1.) -- (2.,1.) -- (-1.,-2.) -- cycle;
\draw [<->] (3,0.) -- (-3,0.);
\draw [<->] (0.,3) -- (0.,-3);
\draw (3,0.) node[anchor=north ] {$\dert z_1$};
\draw (0.,3) node[anchor=west] {$\dert z_2$};
\draw (-3,0.) node[anchor=north ] {$T_1$};
\draw (0.,-3) node[anchor=west] {$T_2$};
\draw [color=red](-1.,1.) node[anchor=south east] {$\alpha_1$};
\draw [color=red](2.,1.) node[anchor=south west] {$\alpha_2$};
\draw [color=red](-1.,-2.)node[anchor=north east] {$\alpha_3$};
\draw (1,-0.5) node[color=red] {$C^*_A$};
\draw [line width=1.2pt,color=red] (-1.,1.)-- (2.,1.);
\draw [line width=1.2pt,color=red] (2.,1.)-- (-1.,-2.);
\draw [line width=1.2pt,color=red] (-1.,-2.)-- (-1.,1.);
\draw [line width=1.pt,->] (0.5,1) -- (0.5,1.7) node[anchor=south] {$\nu_3$};
\draw [line width=1.pt,->] (-1,-0.5) -- (-1.7,-0.5) node[anchor=east] {$\nu_2$};
\draw [line width=1.pt,->] (0.5,-0.5) -- (1,-1) node[anchor=north west] {$\nu_1$};
\end{tikzpicture}
\caption{Case $F_->2F_+$. The stasis domain  $-C_A^*=-\partial\Ds(0)$.}
\label{fig:subdiffA}
\end{figure}

The subdifferential of $\Ds$ in the origin is the convex hull generated by $\alpha_1,\alpha_2,\alpha_3$ (cf.~Fig.~\ref{fig:subdiffA}), namely
\begin{equation}
C_A^*=\partial \Ds(0)=\conv \{\alpha_1,\alpha_2,\alpha_3\}
\end{equation}

If $w\in \inter A_i$, then $\partial \Ds (w)=\alpha_i$, whereas if $w\in A_i\cap A_j\setminus \{0\}$, then $\partial \Ds (w)=\edge{\alpha_i}{\alpha_j}$, where the latter denotes the edge of $C_A^*$ having endpoints $\alpha_i$  and $\alpha_j$, namely $\edge{\alpha_i}{\alpha_j}=\conv\{\alpha_i,\alpha_j\}$. 

\subsection{Case $2F_+>F_->F_+$}

\begin{figure}[h]\centering
\begin{tikzpicture}[line cap=round,line join=round,x=1.0cm,y=1.0cm, scale=0.5]
\clip(-10.,-10.) rectangle (10.,10.);
\fill[line width=1.6pt,color=red,fill=red,fill opacity=0.5] (-1.,-1.333) -- (-0.333,-1.333) -- (1.333,0.333) -- (1.333,1.) -- (-0.333,1.) -- (-1.,0.333) -- cycle;
\draw (4,4.) node {$B_3$};
\draw (-3.,6.) node {$B_2$};
\draw (-7.,3) node {$B_1$};
\draw (-5,-4) node {$B_6$};
\draw (2.,-6.) node {$B_5$};
\draw (7,-3) node {$B_4$};
\draw [line width=1.2pt,color=red] (-1.,-1.333)-- (-0.333,-1.333);
\draw [line width=1.2pt,color=red] (-0.333,-1.333)-- (1.333,0.333);
\draw [line width=1.2pt,color=red] (1.333,0.333)-- (1.333,1.);
\draw [line width=1.2pt,color=red] (1.333,1.)-- (-0.333,1.);
\draw [line width=1.2pt,color=red] (-0.333,1.)-- (-1.,0.333);
\draw [line width=1.2pt,color=red] (-1.,0.333)-- (-1.,-1.333);
\draw [dashed] (-1.,0.)-- (0.,-0.75);
\draw [dashed] (0.,-1.5)-- (-2.,0.);
\draw [dashed] (0.,-2.25)-- (-3.,0.);
\draw [dashed] (-4.,0.)-- (0.,-3.);
\draw [dashed] (0.,-3.75)-- (-5.,0.);
\draw [dashed] (-6.,0.)-- (0.,-4.5);
\draw [dashed] (0.,-0.75)-- (1.,-1.);
\draw [dashed] (2.,-2.)-- (0.,-1.5);
\draw [dashed] (3.,-3.)-- (0.,-2.25);
\draw [dashed] (0.,-3.)-- (4.,-4.);
\draw [dashed] (5.,-5.)-- (0.,-3.75);
\draw [dashed] (0.,-4.5)-- (6.,-6.);
\draw [dashed] (0.75,0.)-- (1.,-1.);
\draw [dashed] (1.5,0.)-- (2.,-2.);
\draw [dashed] (2.25,0.)-- (3.,-3.);
\draw [dashed] (3.,0.)-- (4.,-4.);
\draw [dashed] (3.75,0.)-- (5.,-5.);
\draw [dashed] (4.5,0.)-- (6.,-6.);
\draw [dashed] (0.,1.)-- (0.75,0.);
\draw [dashed] (0.,2.)-- (1.5,0.);
\draw [dashed] (0.,3.)-- (2.25,0.);
\draw [dashed] (0.,4.)-- (3.,0.);
\draw [dashed] (0.,5.)-- (3.75,0.);
\draw [dashed] (0.,6.)-- (4.5,0.);
\draw [dashed] (0.,1.)-- (-0.75,0.75);
\draw [dashed] (0.,2.)-- (-1.5,1.5);
\draw [dashed] (0.,3.)-- (-2.25,2.25);
\draw [dashed] (0.,4.)-- (-3.,3.);
\draw [dashed] (0.,5.)-- (-3.75,3.75);
\draw [dashed] (0.,6.)-- (-4.5,4.5);
\draw [dashed] (-0.75,0.75)-- (-1.,0.);
\draw [dashed] (-1.5,1.5)-- (-2.,0.);
\draw [dashed] (-3.,0.)-- (-2.25,2.25);
\draw [dashed] (-3.,3.)-- (-4.,0.);
\draw [dashed] (-5.,0.)-- (-3.75,3.75);
\draw [dashed] (-4.5,4.5)-- (-6.,0.);
\draw[->,color=black] (-8.,0.) -- (8.,0.);
\draw[->,color=black] (0.,-8.) -- (0.,8.);
\draw [line width=1.2pt,color=Green] (0.,-7.) -- (0.,7.);
\draw [line width=1.2pt,color=Green] (7.,0.) -- (-7.,0.);
\draw [line width=1.2pt,color=Green] (-7.,7.) -- (7.,-7.);
\draw (9,0) node[anchor=north] {$w_1$};
\draw (0.,9.) node[anchor= west] {$w_2$};
\draw (2.,-1.) node[color=red, fill=white] {$C^*_B$};
\end{tikzpicture}
\caption{Case $2F_+>F_->F_+$. The six regions $B_1$,\dots $B_6$, the contour lines of $\Ds$ (dashed) and its subdifferential at the origin $C^*_B$ (red).}
\label{fig:regB}
\end{figure}
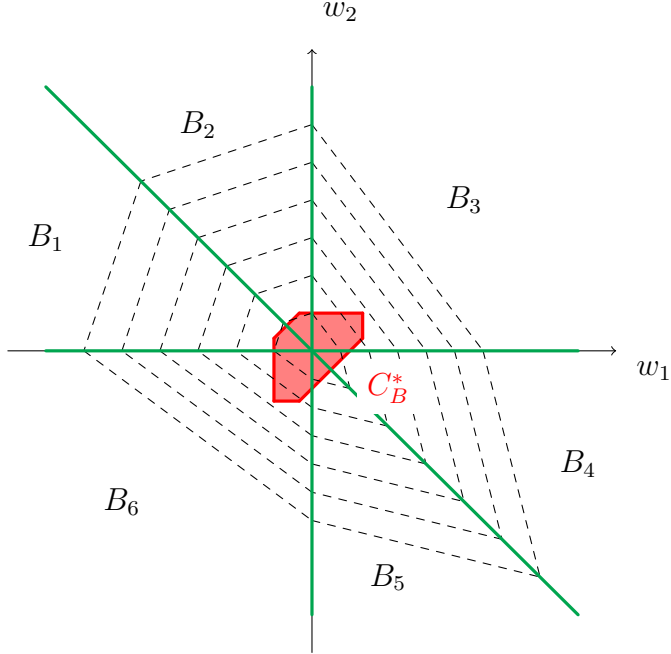

In this case we have to divide the plane into six different regions, as shown in Figure \ref{fig:regB}.
\begin{enumerate}[label=\textup{($B_{\arabic*}$)}]
\item Here $w_1 \leq -w_2 \leq 0$ and so $\vm(w)=-w_2$. In this region we have
\begin{equation*}
\Ds(w)=(-w_1-w_2)F_+ +(w_2)F_-=\scal{\beta_6}{w} \qquad \text{where $\beta_1=\begin{pmatrix}
-F_+  \\ -F_+   +F_-
\end{pmatrix}$}
\end{equation*}

\item Here  $-w_2 \leq w_1 \leq 0$ holds, so $\vm(w)=w_1$. In this region we have
\begin{equation*}
\Ds(w)=(w_1+w_2)F_+ +(-w_1)F_-=\scal{\beta_4}{w} \qquad \text{where $\beta_2=\begin{pmatrix}
F_+  -F_- \\ F_+  
\end{pmatrix}$}
\end{equation*}

\item Here  $-w_2 \leq 0 \leq w_1$ holds, so $\vm(w)=0$. In this region we have
\begin{equation*}
\Ds(w)=(w_2)F_+ +(w_1)F_-=\scal{\beta_2}{w} \qquad \text{where $\beta_3=\begin{pmatrix}
F_- \\ F_+ 
\end{pmatrix}$}
\end{equation*}

\item Here  $0 \leq -w_2 \leq w_1$ holds, so $\vm(w)=-w_2$. In this region we have
\begin{equation*}
\Ds(w)=(-w_2)F_+ +(w_1+w_2)F_-=\scal{\beta_5}{w} \qquad \text{where $\beta_4=\begin{pmatrix}
F_- \\ -F_+   +F_-
\end{pmatrix}$}
\end{equation*}

\item Here  $0 \leq w_1 \leq -w_2$ holds, so $\vm(w)=w_1$. In this region we have
\begin{equation*}
\Ds(w)=(w_1)F_+ +(-w_1-w_2)F_-=\scal{\beta_3}{w} \qquad \text{where $\beta_5=\begin{pmatrix}
F_+ - F_- \\ -F_-
\end{pmatrix}$}
\end{equation*}

\item Here  $w_1 \leq 0 \leq -w_2$ holds, so $\vm(w)=0$. In this region we have
\begin{equation*}
\Ds(w)=(-w_1)F_+ +(-w_2)F_- =\scal{\beta_1}{w} \qquad \text{where $\beta_6=\begin{pmatrix}
-F_+   \\ -F_-
\end{pmatrix}$}
\end{equation*}

\end{enumerate}

The subdifferential of $\Ds$ in the origin is 
\begin{equation}
C_B^*=\partial \Ds(0)=\conv \{\beta_1,\beta_2,\beta_3,\beta_4,\beta_5,\beta_6\}
\end{equation}
If $w\in \inter B_i$, then $\partial \Ds (w)=\beta_i$, whereas if $w\in B_i\cap B_j\setminus \{0\}$, then $\partial \Ds (w)=\edge{\beta_i}{\beta_j}$, using the notation we introduced in the previous case. 

\section{Stasis domains and the laws of motion} \label{sec:law}

We observe that the gradient of $\EE$ with respect to the $z$-coordinates corresponds to the vector composed by the tensions of the two springs, i.e.\
\begin{equation}
D_z\EE(t,z(t))=Az(t)-\ell (t)=\pvector{\frac{k_1}{L_1}\left(z_1(t)-\epsilon_1(t) L_1\right)\\
\frac{k_2}{L_2}\left(z_2(t)-\epsilon_2(t) L_2\right)}=\pvector{T_1(t)\\T_2(t)}=T(t)
\label{eq:defT}
\end{equation}

Thus, from \eqref{eq:SF}, we have
\begin{equation}
-T(t)\in \partial \Ds(\dert z(t))
\label{eq:subdiff}
\end{equation}

We can distinguish between three different situations.
\begin{itemize}
\item If $\dert z(t)=0$, then $-T(t)\in C^*$.
\item If $\dert z(t)\in \inter A_i$ for some $i$, then $-T(t)=\alpha_i$. Similarly, if $\dert z(t)\in \inter B_i$ for some $i$, then $-T(t)=\beta_i$.
\item If $\dert z(t)\in A_i\cap A_j\setminus \{0\}$ for some $i\neq j$, then $-T(t)\in \edge{\alpha_i}{\alpha_j}$. Similarly, if $\dert z(t)\in B_i\cap B_j\setminus \{0\}$ for some $i\neq j$, then $-T(t)\in \edge{\beta_i}{\beta_j}$.
\end{itemize}

This gives us a first description of the behaviour of our system.  The tensions of the springs are allowed to change only within the set $-C^*$, that we call \emph{stasis domain}, in analogy with the elastic domains used in elasto-plasticity. 
Shape changes, and therefore motion, can occur only if the tensions have values on the boundary of $-C^*$, to which we refer as \emph{slip surface}. 

\medskip

The next step is to use the information contained in \eqref{eq:subdiff}, combined with the definition of $T(t)$, to recover how variations in the active distortion produce shape changes. The best way to do that is to work in terms of the tension state of the crawler $T(t)$ instead of the shape state $z(t)$.

First of all we notice that, by differentiating \eqref{eq:defT}, we get
\begin{subequations} \label{eq:derT}
\begin{align}
\dert T_1(t)&=-k_1 \dert \epsilon_1(t)+\frac{k_1}{L_1}\dert z_1(t) \label{eq:derT1}\\
 \dert T_2(t)&=-k_2 \dert \epsilon_2(t)+\frac{k_2}{L_2}\dert z_2(t) \label{eq:derT2}
\end{align}
\end{subequations}
If $-T(t)\in \inter C^*$, from \eqref{eq:subdiff} we have $\dert z(t)=0$ and the previous equations reduce to
\begin{equation}
\dert T_1(t)=-k_1 \dert \epsilon_1(t) \qquad \dert T_2(t)=-k_2 \dert \epsilon_2(t)
\label{eq:mot_intC*}
\end{equation}
that describe the evolution of the system. On the other end, when $T(t)$ lies on the boundary of $-C^*$, the behaviour of the system is less trivial. We will discuss first the simpler case $F_->2F_+$ and then consider the second case $2F_+>F_->F_+$.


\subsection{Case $F_->2F_+$}

First of all let us introduce the unit vectors
\begin{align*}
 \nu_1=\frac{1}{\sqrt{2}}\pvector{1\\-1} && \nu_2 =\pvector{-1\\0} && \nu_3=\pvector{0\\1}
\end{align*}
that are the outer unit normals to $C^*_A$ respectively along the edges $\edge{\alpha_2}{\alpha_3}$, $\edge{\alpha_3}{\alpha_1}$ and $\edge{\alpha_1}{\alpha_2}$.
The constraint $-T(t)\in C^*_A$ implies that, if $T$ is differentiable at time $t$, then
\begin{align}
\scal{\dert T(t)}{\nu_1}= 0  &&\text{if $-T(t)\in \edge{\alpha_2}{\alpha_3}$}\notag \\
\scal{\dert T(t)}{\nu_2}= 0  &&\text{if $-T(t)\in \edge{\alpha_3}{\alpha_1}$}\label{eq:bound_ineq} \\
\scal{\dert T(t)}{\nu_3}= 0  &&\text{if $-T(t)\in \edge{\alpha_1}{\alpha_2}$}\notag
\end{align}
If one of the scalar products were positive, then the tension should have been outside the stasis domain $C^*_A$ for the times immediately before, and similarly, if one of them were negative, the tension would be outside $C^*_A$ for the times immediately after. 

Let us note that condition \ref{eq:bound_ineq} can be expressed in a more concise way as
\begin{equation}
-T(t)\in N_{C^*_A}(T(t))
\end{equation}
where $N_C(T)$ denotes the normal cone to the convex set $C$ at the point $T$. This is also a classical way to approach the problem \eqref{eq:RVI}, usually known as \emph{differential inclusion} formulation \cite{Mie04, Mie05}. 

Following this same line of thought, each of the constraints could be decoupled into two inequalities on the increments of $T$, one for the past and one for the future, without requiring the differentiability of $T$.
However, for our purposes, we will work under the assumptions of Theorem \ref{th:mielke}, that guarantees  the Lipschitz continuity of the tension $T(t)$, so that the times when $T(t)$ is not differentiable can be neglected for the study of the motion.

A consequence of \eqref{eq:bound_ineq} is that, when we reach an edge, either the tension is differentiable, that implies $\scal{\dert \ell(t)}{\nu_i}=0$  and thus means that $\epsilon(t)$ is in a certain sense \lq\lq well calibrated\rq\rq , or we have a time $t$ of non-differentiability for $T(t)$ and $z(t)$, corresponding to an abrupt transition between rest and motion.

If $-T(t)$ lies on one on the vertices of $C^*_A$, then two of the constraints of \eqref{eq:bound_ineq} are satisfied simultaneously, leading to
\begin{align}
\dert z_1(t)=L_1\dert \epsilon_1(t) &&
\dert z_2(t) =L_2\dert \epsilon_2(t)
\label{eq:mot_Avert}
\end{align}
We also recall that, by \eqref{eq:subdiff}, we know that $\dert z(t)\in A_i$; combining this with \eqref{eq:mot_Avert} we see that, to keep that tension configuration, the derivative of the active distortion must lie in a specific cone. In more detail, we have the following situation.

\begin{itemize}
\item If $-T(t)=\alpha_1$, then by \eqref{eq:bound_ineq} we have $\dert z_1(t)\leq 0\leq \dert z_2(t)$ (i.e. $\dert z(t)\in A_1$), that implies $\dert v(t)=0$ and 
\begin{equation*}
\dert \epsilon_1(t)\leq 0\leq \dert \epsilon_2(t)
\end{equation*}
The resulting motion of the crawler is
\begin{align*}
\dert u_1(t)=-L_1 \dert \epsilon_1(t)\geq 0
&& \dert u_2(t)=0
&& \dert u_3(t)=L_2 \dert \epsilon_2(t)\geq 0
\end{align*}
\item If $-T(t)=\alpha_2$, then  $\dert z_1(t)\geq 0$ and $\dert z_1(t)\geq -\dert z_2(t)$, so that $\dert v(t)=\dert z_1(t)$ and 
\begin{equation*}
\dert \epsilon_1(t)\geq 0 \qquad \text{and}\qquad \dert \epsilon_2(t)\geq - \frac{L_1}{L_2}\dert \epsilon_1(t)
\end{equation*}
The resulting motion of the crawler is
\begin{align*}
&\dert u_1(t)=0
& \dert u_2(t)=L_1 \dert \epsilon_1(t)\geq 0\\
&\dert u_3(t)=L_1 \dert \epsilon_1(t)+L_2 \dert \epsilon_2(t) \geq 0
\end{align*}
\item If $-T(t)=\alpha_3$, then  $\dert z_2(t)\leq 0$ and $\dert z_1(t)\leq -\dert z_2(t)$, so that $\dert v(t)=-\dert z_2(t)$ and 
\begin{equation*}
\dert \epsilon_2(t)\leq 0 \qquad \text{and}\qquad \dert \epsilon_1(t)\leq - \frac{L_2}{L_1}\dert \epsilon_2(t)
\end{equation*}
The resulting motion of the crawler is
\begin{align*}
&\dert u_1(t)=-L_1 \dert \epsilon_1(t)-L_2 \dert \epsilon_2(t) \geq 0
& \dert u_2(t)=-L_2 \dert \epsilon_2(t)\geq 0\\
&\dert u_3(t)=0
\end{align*}
\end{itemize}

If $-T(t)$ lies in the interior of one on the edges of $C^*_A$, then condition \eqref{eq:bound_ineq} gives us only one constraint. However a second constraint is obtained by \eqref{eq:subdiff}, since we know that, if $-T(t)\in\inter \edge{\alpha_i}{\alpha_j}$, then $\dert z(t)\in A_i\cap A_j$.  
In more detail, we have the following situation.
\begin{itemize}
\item If $-T(t)\in\edge{\alpha_1}{\alpha_2}$ then we have $\dert v(t)=0$ and
\begin{align*}
&\dert z_1(t)=0 &
&\dert z_2(t) =L_2\dert \epsilon_2(t)\geq 0\\
&\dert T_1(t)=-k_1 \dert \epsilon_1(t) &
&\dert T_2(t)=0
\end{align*}
The resulting motion of the crawler is
\begin{align*}
\dert u_1(t)=\dert u_2(t)=0
&&\dert u_3(t)=L_2 \dert \epsilon_2(t)\geq 0
\end{align*}
\item If $-T(t)\in\edge{\alpha_3}{\alpha_1}$ then we have $\dert v(t)=0$ and
\begin{align*}
&\dert z_1(t)=-L_1\dert \epsilon_1(t)\geq 0 &
&\dert z_2(t) =0\\
&\dert T_1(t)=0 &
&\dert T_2(t)=-k_2 \dert \epsilon_2(t)
\end{align*}
The resulting motion of the crawler is
\begin{align*}
\dert u_1(t)=-L_1 \dert \epsilon_1(t)\geq 0
&&\dert u_2(t)=\dert u_3(t)=0
\end{align*}
\item If $-T(t)\in\edge{\alpha_2}{\alpha_3}$, differently from the two previous cases, we observe changes on the tension and length of both segments; however this happens in a coordinated fashion, namely,
\begin{equation*}
\dert z_1(t)=-\dert z_2(t)=\dert v(t)=\frac{k_1\dert \epsilon_1(t)-k_2\dert \epsilon_2(t)}{\frac{k_1}{L_1}+\frac{k_2}{L_2}}\geq 0 
\end{equation*}
that gives the condition $\dert \epsilon_1(t)\geq \frac{k_2}{k_1}\dert \epsilon_2(t)$ for the admissible active distortion. The tension  evolves according to
\begin{equation*}
\dert T_1(t)=\dert T_2(t)=-\frac{L_1\dert \epsilon_1(t)+L_2\dert \epsilon_2(t)}{\frac{L_1}{k_1}+\frac{L_2}{k_2}}
\end{equation*}
The resulting motion of the crawler is
\begin{align*}
\dert u_1(t)=\dert u_3(t)=0
&&\dert u_2(t)=\frac{k_1\dert \epsilon_1(t)-k_2\dert \epsilon_2(t)}{\frac{k_1}{L_1}+\frac{k_2}{L_2}}\geq 0 
\end{align*}
\end{itemize}


\subsection{Case $2F_+>F_->F_+$}

\begin{figure}[tb]\centering
\begin{tikzpicture}[line cap=round,line join=round,x=1.cm,y=1.0cm,>=stealth,scale=1.2]
\clip(-4.,-4.) rectangle (4.,4.);
\fill[line width=1.2pt,color=red,fill=red,fill opacity=0.5] (-1.,-1.333) -- (-0.3333,-1.333) -- (1.333,0.3333) -- (1.333,1.) -- (-0.3333,1.) -- (-1.,0.3333) -- cycle;
\draw [line width=1.2pt,color=red] (-1.,-1.333)-- (-0.3333,-1.333) node[anchor=north ] {$\beta_5$};
\draw [line width=1.2pt,color=red] (-0.3333,-1.333)-- (1.333,0.3333)  node[anchor= west] {$\beta_4$};
\draw [line width=1.2pt,color=red] (1.333,0.3333)-- (1.333,1.) node[anchor=south west] {$\beta_3$};
\draw [line width=1.2pt,color=red] (1.333,1.)-- (-0.3333,1.) node[anchor=south east] {$\beta_2$};
\draw [line width=1.2pt,color=red] (-0.3333,1.)-- (-1.,0.3333) node[anchor=south east] {$\beta_1$};
\draw [line width=1.2pt,color=red] (-1.,0.3333)-- (-1.,-1.333) node[anchor=north east] {$\beta_6$};
\draw [<->] (3,0.) -- (-3,0.);
\draw [<->] (0.,3) -- (0.,-3);
\draw (3,0.) node[anchor=north ] {$\dert z_1$};
\draw (0.,3) node[anchor=west] {$\dert z_2$};
\draw (-3,0.) node[anchor=north ] {$T_1$};
\draw (0.,-3) node[anchor=west] {$T_2$};
\draw (1,-0.5) node[color=red] {$C^*_B$};
\draw [line width=1.pt,->] (0.5,1) -- (0.5,1.7) node[anchor=south] {$\nu_3$};
\draw [line width=1.pt,->] (-1,-0.5) -- (-1.7,-0.5) node[anchor=east] {$\nu_2$};
\draw [line width=1.pt,->] (0.5,-0.5) -- (1,-1) node[anchor=north west] {$\nu_1$};
\draw [line width=1.pt,->] (-0.6666,-1.333) -- (-0.6666,-2.033) node[anchor=north] {$-\nu_3$};
\draw [line width=1.pt,->] (1.333,0.6666) -- (2.033,0.6666) node[anchor=west] {$-\nu_2$};
\draw [line width=1.pt,->] (-0.6666,0.6666) -- (-1.1666,1.1666) node[anchor=south east] {$-\nu_1$};
\end{tikzpicture}
\caption{Case $2F_+>F_->F_+$. The stasis domain  $-C_B^*=-\partial\Ds(0)$.}
\label{fig:subdiffB}
\end{figure}
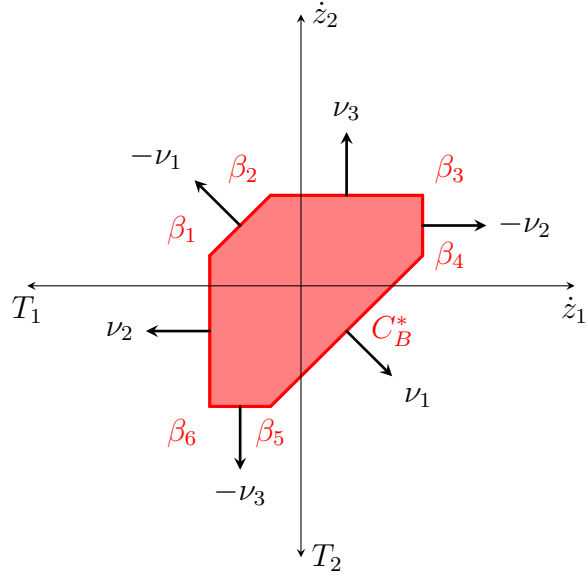

As in the previous case, we want to exploit the constraint $-T(t)\in C^*_B$ to deduce a condition on $\dert T(t)$. We observe that $\nu_1$, $\nu_2$ and $\nu_3$ are the outer unit normals respectively to the edges $\edge{\beta_4}{\beta_5}$, $\edge{\beta_6}{\beta_1}$ and $\edge{\beta_2}{\beta_3}$, but also the inner unit normals to the edges $\edge{\beta_1}{\beta_2}$, $\edge{\beta_3}{\beta_4}$ and $\edge{\beta_5}{\beta_6}$. Thus we have, analogously to \eqref{eq:bound_ineq},
\begin{align}
\scal{\dert T(t)}{\nu_1}= 0  &&\text{if $-T(t)\in \edge{\beta_4}{\beta_5}\cup \edge{\beta_1}{\beta_2}$}\notag \\
\scal{\dert T(t)}{\nu_2}= 0  &&\text{if $-T(t)\in \edge{\beta_6}{\beta_1}\cup \edge{\beta_3}{\beta_4}$}\label{eq:bound_ineqB} \\
\scal{\dert T(t)}{\nu_3}= 0  &&\text{if $-T(t)\in \edge{\beta_2}{\beta_3}\cup \edge{\beta_5}{\beta_6}$}\notag
\end{align}

As before, when $-T(t)$ lies in one on the vertices of $C^*_B$, two of the constraints of \eqref{eq:bound_ineqB} are satisfied simultaneously and therefore
\begin{align}
\dert z_1(t)=L_1\dert \epsilon_1(t) &&
\dert z_2(t) =L_2\dert \epsilon_2(t)
\label{eq:mot_AvertB}
\end{align}
Similarly to the previous case, if $-T(t)\in \beta_i$, then by \eqref{eq:subdiff} we have $\dert z(t)\in B_i$, leading to the following situation.
\begin{itemize}
\item If $-T(t)=\beta_1$, then by \eqref{eq:bound_ineq} we have $\dert z_1(t)\leq -\dert z_2(t) \leq 0 $, that implies $\dert v(t)=-\dert z_2(t)$ and requires, when $T(t)$ is differentiable, that
\begin{align*}
\dert \epsilon_2(t)\geq 0   &&
\dert \epsilon_1(t)\leq -\frac{L_2}{L_1}\dert \epsilon_2(t)
\end{align*}
The resulting motion of the crawler is
\begin{align*}
\dert u_1(t)=-L_1 \dert \epsilon_1(t)-L_2 \dert \epsilon_2(t)\geq 0
&& \dert u_2(t)=-L_2 \dert \epsilon_2(t)\leq 0
&& \dert u_3(t)=0
\end{align*}
\item If $-T(t)=\beta_2$, then  we have $-\dert z_2(t)\leq \dert z_1(t) \leq 0 $, so that $\dert v(t)=\dert z_1(t)$ and 
\begin{align*}
\dert \epsilon_1(t)\leq 0   &&
\dert \epsilon_2(t)\geq -\frac{L_1}{L_2}\dert \epsilon_1(t)
\end{align*}
The resulting motion of the crawler is
\begin{align*}
\dert u_1(t)=0
&& \dert u_2(t)=L_1 \dert \epsilon_1(t)\leq 0
&& \dert u_3(t)=L_1 \dert \epsilon_1(t)+L_2 \dert \epsilon_2(t)\geq 0
\end{align*}
\item If $-T(t)=\beta_3$, then  we have $\dert z_1(t)\geq 0 $ and $\dert z_2(t)\geq 0$,  so that $\dert v(t)=0$ and 
\begin{align*}
\dert \epsilon_1(t)\geq 0   &&
\dert \epsilon_2(t)\geq 0
\end{align*}
The resulting motion of the crawler is
\begin{align*}
\dert u_1(t)=-L_1 \dert \epsilon_1(t)\leq 0
&& \dert u_2(t)=0
&& \dert u_3(t)=L_2 \dert \epsilon_2(t)\geq 0
\end{align*}
\item If $-T(t)=\beta_4$, then by \eqref{eq:bound_ineq} we have $\dert z_1(t)\geq -\dert z_2(t) \geq 0 $, so that $\dert v(t)=-\dert z_2(t)$ and 
\begin{align*}
\dert \epsilon_2(t)\leq 0   &&
\dert \epsilon_1(t)\geq -\frac{L_2}{L_1}\dert \epsilon_2(t)
\end{align*}
The resulting motion of the crawler is
\begin{align*}
\dert u_1(t)=-L_1 \dert \epsilon_1(t)-L_2 \dert \epsilon_2(t)\leq 0
&& \dert u_2(t)=-L_2 \dert \epsilon_2(t)\geq 0
&& \dert u_3(t)=0
\end{align*}
\item If $-T(t)=\beta_5$, then  we have $-\dert z_2(t)\geq \dert z_1(t) \geq 0 $, so that $\dert v(t)=\dert z_1(t)$ and 
\begin{align*}
\dert \epsilon_1(t)\geq 0   &&
\dert \epsilon_2(t)\leq -\frac{L_1}{L_2}\dert \epsilon_1(t)
\end{align*}
The resulting motion of the crawler is
\begin{align*}
\dert u_1(t)=0
&& \dert u_2(t)=L_1 \dert \epsilon_1(t)\geq 0
&& \dert u_3(t)=L_1 \dert \epsilon_1(t)+L_2 \dert \epsilon_2(t)\leq 0
\end{align*}
\item If $-T(t)=\beta_6$, then  we have $\dert z_1(t)\leq 0 $ and $\dert z_2(t)\leq 0$,  so that $\dert v(t)=0$ and 
\begin{align*}
\dert \epsilon_1(t)\leq 0   &&
\dert \epsilon_2(t)\leq 0
\end{align*}
The resulting motion of the crawler is
\begin{align*}
\dert u_1(t)=-L_1 \dert \epsilon_1(t)\geq 0
&& \dert u_2(t)=0
&& \dert u_3(t)=L_2 \dert \epsilon_2(t)\leq 0
\end{align*}
\end{itemize}

As in the previous case, when $-T(t)$ lies in the interior of one on the edges of $C^*_B$, only one constraint is given by condition \eqref{eq:bound_ineqB}, but a second one is recovered by \eqref{eq:subdiff}, using the fact that if $-T(t)\in\inter \edge{\beta_i}{\beta_j}$, then $\dert z(t)\in B_i\cap B_j$.  The pairs of opposite edges are characterized by the same behaviour of the crawler, but associated with shape variations of opposite sign.
In more detail, we have the following situation.

\begin{itemize}
\item If $-T(t)\in\edge{\beta_2}{\beta_3}\cup \edge{\beta_5}{\beta_6}$ then we have $\dert v(t)=0$ and
\begin{align*}
&\dert z_1(t)=0 &
&\dert z_2(t) =L_2\dert \epsilon_2(t)\\
&\dert T_1(t)=-k_1 \dert \epsilon_1(t) &
&\dert T_2(t)=0
\end{align*}
so that it is required that $\epsilon_2(t)\geq 0$ if $-T(t)\in\edge{\beta_2}{\beta_3}$, whereas $\epsilon_2(t)\leq 0$ if $-T(t)\in\edge{\beta_5}{\beta_6}$.
The resulting motion of the crawler is
\begin{align*}
\dert u_1(t)=\dert u_2(t)=0
&&\dert u_3(t)=L_2 \dert \epsilon_2(t)\begin{cases}
\geq 0 & \text{if $-T(t)\in\edge{\beta_2}{\beta_3}$}\\ \leq 0 & \text{if $-T(t)\in\edge{\beta_5}{\beta_6}$}
\end{cases}
\end{align*}
\item If $-T(t)\in\edge{\beta_3}{\beta_4}\cup \edge{\beta_6}{\beta_1}$ then we have $\dert v(t)=0$ and
\begin{align*}
&\dert z_1(t)=-L_1\dert \epsilon_1(t) &
&\dert z_2(t) =0\\
&\dert T_1(t)=0 &
&\dert T_2(t)=-k_2 \dert \epsilon_2(t)
\end{align*}
so that it is required that $\epsilon_2(t)\geq 0$ if $-T(t)\in\edge{\beta_6}{\beta_1}$, whereas $\epsilon_2(t)\leq 0$ if $-T(t)\in\edge{\beta_3}{\beta_4}$.
The resulting motion of the crawler is
\begin{align*}
\dert u_2(t)=\dert u_3(t)=0
&&\dert u_1(t)=-L_1 \dert \epsilon_1(t)
\begin{cases}
\geq 0 & \text{if $-T(t)\in\edge{\beta_6}{\beta_1}$}\\ \leq 0 & \text{if $-T(t)\in\edge{\beta_3}{\beta_4}$}
\end{cases}
\end{align*}
\item The third case $-T(t)\in\edge{\beta_1}{\beta_2}\cup \edge{\beta_4}{\beta_5}$, is characterized by a coordinated change in the tension and length of both segments, more precisely
\begin{equation*}
\dert z_1(t)=-\dert z_2(t)=\dert v(t)=\frac{k_1\dert \epsilon_1(t)-k_2\dert \epsilon_2(t)}{\frac{k_1}{L_1}+\frac{k_2}{L_2}}
\end{equation*}
that gives, for the admissible active distortion, the condition $\dert \epsilon_1(t)\geq \frac{k_2}{k_1}\dert \epsilon_2(t)$ if $-T(t)\in\edge{\beta_4}{\beta_5}$ and  $\dert \epsilon_1(t)\leq \frac{k_2}{k_1}\dert \epsilon_2(t)$ if $-T(t)\in\edge{\beta_1}{\beta_2}$. The tension configuration evolves according to
\begin{equation*}
\dert T_1(t)=\dert T_2(t)=-\frac{L_1\dert \epsilon_1(t)+L_2\dert \epsilon_2(t)}{\frac{L_1}{k_1}+\frac{L_2}{k_2}}
\end{equation*}
The resulting motion of the crawler is
\begin{align*}
\dert u_1(t)=\dert u_3(t)=0
&&\dert u_2(t)=\frac{k_1\dert \epsilon_1(t)-k_2\dert \epsilon_2(t)}{\frac{k_1}{L_1}+\frac{k_2}{L_2}}
\begin{cases}
\geq 0 & \text{if $-T(t)\in\edge{\beta_4}{\beta_5}$}\\ \leq 0 & \text{if $-T(t)\in\edge{\beta_1}{\beta_2}$}
\end{cases}
\end{align*}
\end{itemize}



\section{Motility analysis and crawling strategies} \label{sec:disc}

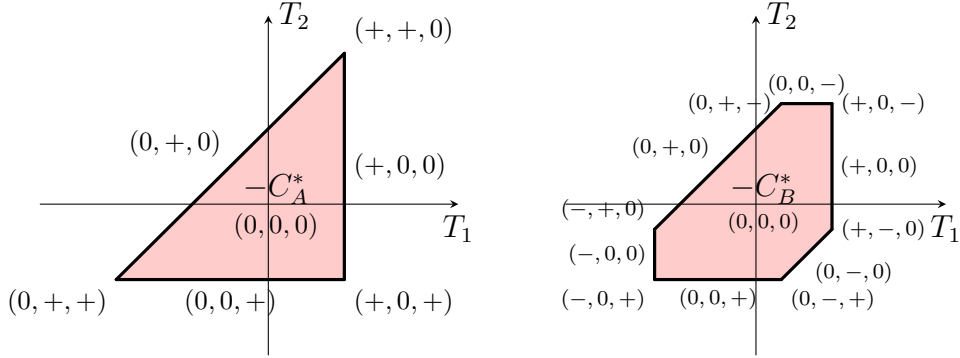
\begin{figure}[tb]
\centering
\subfloat[][\emph{Case $F_- > 2F_+$}.]
{\begin{tikzpicture}[line cap=round,line join=round,x=1.0cm,y=1.0cm,>=stealth,scale=1.]
\clip(-3.5,-3.) rectangle (3.,3.);
\fill[line width=1 pt,color=red,fill=red,fill opacity=0.2] (1.,-1.) -- (-2.,-1.) -- (1.,2.) -- cycle;
\draw [<-] (2.5,0.) -- (-3,0.);
\draw [<-] (0.,2.5) -- (0.,-2);
\draw (2.5,0.) node[anchor=north ] {$T_1$};
\draw (0.,2.5) node[anchor=west] {$T_2$};
\draw (0.1,0.2) node {$-C^*_A$}; \footnotesize
\draw [](1.,-1.) node[anchor=north west] {$(+,0,+)$};
\draw [](-2.,-1.) node[anchor=north east] {$(0,+,+)$};
\draw [](1.,2.)node[anchor=south west] {$(+,+,0)$};
\draw [] (-0.5,-1) node[anchor=north] {$(0,0,+)$};
\draw [](1,0.5) node[anchor= west] {$(+,0,0)$};
\draw [](-0.5,0.5)node[anchor=south east] {$(0,+,0)$};
\draw [](0.1,0)node[anchor=north] {$(0,0,0)$};
\draw [line width=1.2pt] (1.,-1.)-- (-2.,-1.);
\draw [line width=1.2pt] (-2.,-1.)-- (1.,2.);
\draw [line width=1.2pt] (1.,2.)-- (1.,-1.);
\end{tikzpicture}}\quad
\subfloat[][\emph{Case $2F_+> F_- >F_+$}.]
{\begin{tikzpicture}[line cap=round,line join=round,x=1.cm,y=1.0cm,>=stealth,scale=1.]
\clip(-3.,-3.) rectangle (3.,3.);
\fill[line width=1.2pt,color=red,fill=red,fill opacity=0.2] (1.,1.333) -- (0.3333,1.333) -- (-1.333,-0.3333) -- (-1.333,-1.) -- (0.3333,-1.) -- (1.,-0.3333) -- cycle;\scriptsize
\draw [line width=1.2pt] (1.,1.333)-- (0.3333,1.333) node[anchor=east ] {$(0,+,-)$};
\draw [line width=1.2pt] (0.3333,1.333)-- (-1.333,-0.3333)  node[anchor=south east] {$(-,+,0)$};
\draw [line width=1.2pt] (-1.333,-0.3333)-- (-1.333,-1.) node[anchor=north east] {$(-,0,+)$};
\draw [line width=1.2pt] (-1.333,-1.)-- (0.3333,-1.) node[anchor=north west] {$(0,-,+)$};
\draw [line width=1.2pt] (0.3333,-1.)-- (1.,-0.3333) node[anchor= west] {$(+,-,0)$};
\draw [line width=1.2pt] (1.,-0.3333)-- (1.,1.333) node[anchor= west] {$(+,0,-)$};
\draw [<-] (2.5,0.) -- (-2.5,0.);
\draw [<-] (0.,2.5) -- (0.,-2);\normalsize
\draw (2.5,0.) node[anchor=north ] {$T_1$};
\draw (0.,2.5) node[anchor=west] {$T_2$};
\draw (0.1,0.2) node[] {$-C^*_B$}; \scriptsize
\draw  (-0.5,-1) node[anchor=north] {$(0,0,+)$};
\draw  (1,0.5) node[anchor=west] {$(+,0,0)$};
\draw  (-0.5,0.5)  node[anchor=south east] {$(0,+,0)$};
\draw  (0.6666,1.333) node[anchor=south] {$(0,0,-)$};
\draw  (-1.333,-0.6666)  node[anchor=east] {$(-,0,0)$};
\draw (0.6666,-0.6666)  node[anchor=north west] {$(0,-,0)$};
\draw [](0.1,0)node[anchor=north] {$(0,0,0)$};
\end{tikzpicture}} 
\caption{Qualitative summary of the motility results of section \ref{sec:law}. Each triple is placed in the interior, on an edge or on a vertex of the stasis domain $-C^*$ and describes the admissible directions of displacement for the three legs while the crawler keeps that tension configuration. A plus denotes a positive displacement, a minus a negative one and a zero that that leg must remain steady. 
For instance the triple $(+,0,-)$ near a vertex indicates that, for that value of the tension $T(t)$, we have $\dert u_1(t)\geq 0$, $\dert u_2(t)=0$ and $\dert u_3(t)\leq 0$.}
\label{fig:mot}
\end{figure}

A qualitative description of the results of the previous section is illustrated in Figure \ref{fig:mot}. The two possibilities considered for the relative magnitude of the friction forces determine very different motile behaviours of the crawler.

If $F_->2F_+$, the legs of the crawler can move only forward. The set $-C_A^*$ of the admissible tension configurations scales with $F_+$, but it is independent of the value of $F_-$.

If $2F_+>F_->F_+$, each leg of the crawler can move both forward and backward. The precise shape of the stasis domain $-C_B^*$ depends on the ratio $F_+/F_-$, although it is always a hexagon with parallel opposite edges oriented as in Figure \ref{fig:subdiffB}. If the ratio $F_+/F_-$ is fixed, then $-C_B^*$ scales homothetically with the magnitude of the friction coefficients; if instead we fix the value of $F_+$, then $-
C_B^*$ shrinks as $F_-$ tends to $F_+$.

To truly understand the motility of our crawler, we have to consider the effects of a periodic active distortion $\epsilon(t)$. As a corollary of Theorem \ref{th:mielke}, we are granted the existence of a unique Lipschitz continuous displacement $u(X,t)$ for any given continuous and piecewise continuously differentiable active distortion $\epsilon\colon [0,\T]\to \R^2$.

\medbreak

We now discuss the main qualitative behaviour of such motility strategies and then present some illustrative examples. To simplify the computation, we assume $k_1=k_2=k$  and $L_1=L_2=L$.

To produce a non-null translation of the crawler that repeats itself in each period, sufficiently large excursions in the stasis domain are necessary. 
More precisely, during every period the tension $T(t)$ has to reach all the three edges of $-C^*_A$ (if $F_->2F_+$) or a suitable triple of non adjacent edges of $-C_B^*$ (if $2F_+>F_->F_+$).
Since a certain amount of excursion in the active distortion is spent in crossing $-C^*$, allowing larger fluctuations in $\epsilon(t)$ permits more performant motility strategies, because in this way a larger amount of the active distortion is spent  moving the legs.

In the case $F_->2F_+$, an effective motility strategy can be achieved even by activating only one of the segments, for instance by setting $\epsilon_2\equiv 0$ and assuming a sufficiently large sawtooth oscillation for $\epsilon_1$. This strategy can be compared to a one-segment crawler experiencing  the same sawtooth fluctuations, as that studied in \cite[Sec.~4]{DeSGidNos15}. Indeed, the one-segment crawler results more efficient: it requires a lower minimal amplitude $\Delta \epsilon$ of the sawtooth ($\Delta\epsilon>2F_+/k$ instead of $\Delta\epsilon>3F_+/k$), it produces a greater displacement after one cycle ($\Delta u=(\Delta \epsilon-2F_+/k)L$ instead of $\Delta u=(\Delta \epsilon-3F_+/k)L$ ) and it is effective also in the case $2F_+>F_->F_+$. For such friction ratios  a two-segment crawler, performing the sawtooth strategy above, has a zero net displacement after one cycle.

We remark that in all the situations above, net  displacements are possible only in the direction of lower friction.
To achieve a \emph{complete motility}, i.e.~to be able to move also backwards (against the higher friction) using periodic shape changes, we need to consider the case $2F_+>F_->F_+$ and strategies that fully exploit two shape parameters. This minimality of two shape parameters for a complete motility belongs to folklore knowledge for unidimensional locomotors (cf.~for instance \cite{Arroyo,DeSTat12,GidNosDeS14,MonDeS15}). The ability of our two-segment crawler to effectively move in both directions, assuming a small friction asymmetry, is illustrated by the following strategies.

\begin{figure}[tb]
\centering
\subfloat[][]
{\begin{tikzpicture}[line cap=round,line join=round,>=stealth,x=1.0cm,y=1.0cm,line width=1.pt,scale=0.8]
\clip(-1,-0.5) rectangle (6.5,6.5);
\draw [->] (0.,0.) -- (6.,0.) node[anchor=north] {$\epsilon_1$};
\draw [->] (0.,0.) -- (0.,6.) node[anchor=east] {$\epsilon_2$};\footnotesize
\draw [->,color=red] (1.,5.) -- (1,3) node [anchor=east] {$\gamma_1$}; 
\draw [->,color=red] (1.,1.) --  (3,1) node [anchor=north] {$\gamma_2$};
\draw [->,color=red] (5.,1.) --  (3,3) node [anchor=south west] {$\gamma_3$};
\draw [->,color=red](1,3)-- (1.,1.);
\draw [->,color=red](3,1)-- (5.,1.);
\draw [->,color=red](3,3)-- (1.,5.);
\end{tikzpicture}}\quad
\subfloat[][]
{\begin{tikzpicture}[line cap=round,line join=round,x=1.cm,y=1.0cm,>=stealth,decoration={
    markings,
    mark=at position 0.5 with {\arrow{>}}},scale=1.6]
\clip(-2.,-1.5) rectangle (2.2,2.2);
\draw[line width=1.pt,color=red,fill=red,fill opacity=0.0] (1.,1.333) -- (0.3333,1.333) -- (-1.333,-0.3333) -- (-1.333,-1.) -- (0.3333,-1.) -- (1.,-0.3333) -- cycle;
\draw [<-] (2.,0.) -- (-2.,0.);
\draw [<-] (0.,2.) -- (0.,-1.5);\normalsize
\draw (2.,0.) node[anchor=north ] {$T_1$};
\draw (0.,2.) node[anchor=west] {$T_2$};
\draw[->,line width=1.2pt, dashed,postaction={decorate}] (0.3333,-1.) --(0.3333,1.333);
\draw[->,line width=1.2pt, dashed,postaction={decorate}] (0.3333,1.333) -- (-1.333,-0.3333);
\draw[->,line width=1.2pt, dashed,postaction={decorate}] (-1.333,-0.3333) -- (-0.666,-1);
\draw[->,line width=1.2pt, dashed,postaction={decorate}] (-0.666,-1) -- (0.3333,-1.);
\draw (0.333,0.333) node[anchor=west] {$\gamma_1$};
\draw (-0.5,0.5) node[anchor=south east] {$\gamma_2$};
\draw (-0.666,-1) node[anchor=south west] {$\gamma_3$};
\end{tikzpicture}} 
\caption{Active distortion strategy \eqref{eq:example} and associated evolution of the tension.}
\label{fig:ex1}
\end{figure}
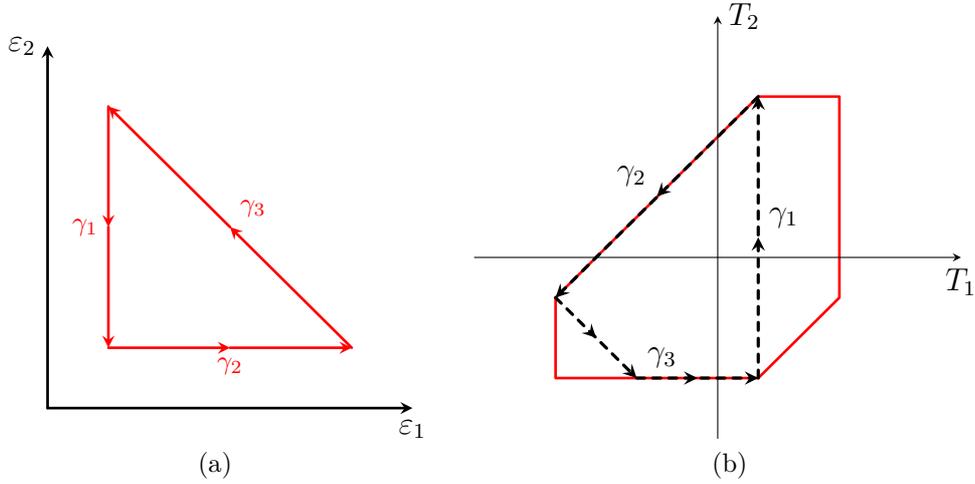

\medbreak

We consider the periodic change in the active distortion illustrated in Figure \ref{fig:ex1}, recalling that $2F_+>F_->F_+$. We set the times so that the period is $\T=3\tau$ and divide the evolution of $\epsilon (t)$ into three phases, described as follows.
\begin{align}
\dert \epsilon_1(t)=\begin{cases}
0 &\text{if $0<t<\tau$}\\
\eta &\text{if $\tau<t<2\tau$}\\
-\eta &\text{if $2\tau<t<3\tau$}\\
\end{cases} &&
\dert \epsilon_2(t)=\begin{cases}
-\eta &\text{if $0<t<\tau$}\\
0 &\text{if $\tau<t<2\tau$}\\
\eta &\text{if $2\tau<t<3\tau$}\\
\end{cases}
\label{eq:example}
\end{align}
where $\eta>0$ is a given parameter. We require that $\eta \tau k>F_++F_-$, to ensure sufficiently large distortions. Note that, since our system is rate independent, what really affects the resulting displacement is not $\eta$ but the increment $\eta \tau$ of the active distortion; actually, any other smooth time reparametrization of the curve in Figure \ref{fig:ex1} (a) would produce exactly the same displacement after each period. 

The behaviour of the system in the first period depends on the initial state; however after the first period we always reach the same tension configuration $T(3\tau)=-\beta_2$. Since we are interested in the long term behaviour, we assume $T(0)=-\beta_2$ and so avoid the initial adjustment period.

We now describe the behaviour in the three phases (see Fig. \ref{fig:ex1}).
\begin{enumerate}[label=\textup{($\gamma_{\arabic*}$)}]
\item For $0<t<\frac{F_++F_-}{\eta k}$ the three legs are steady and $T_2$ increases from $-F_+$ to $F_-$. Then, for $\frac{F_++F_-}{\eta k}<t<\tau$ the tension are constant but the third leg moves backwards with $\dert u_3(t)=-\eta L$.
\item For $\tau<t<\frac{4F_--2F_+}{\eta k}$ the tension evolves from $-\beta_5$ to $-\beta_4$ along the corresponding edge of $-C_B^*$. At the same time, the middle leg moves forward with $\dert u_2(t)=-\frac{\eta L}{2}$. Once the tension edge $-\beta_4$ is reached, for $\frac{4F_--2F_+}{\eta k}<t<2\tau$ the tension is constant, the middle leg is again steady while the first leg moves backwards with  $\dert u_1(t)=-\eta L$.
\item For $2\tau<t<\frac{2F_+-F_-}{k}$, $T_1$ increases and $T_2$ decreases at the same rate, until they reach the edge of  $-C_B^*$. Then, for $\frac{2F_+-F_-}{k}<t<\frac{3F_--3F_+}{k}$ the tension evolves along the edge until it reaches the vertex $-\beta_2$. In this time interval the third leg advances with $\dert u_3(t)=L\eta$. Finally, in the last interval $\frac{3F_--3F_+}{k}<t<3\tau$, the tension is constant, the third leg is again steady and the middle leg moves backwards with $\dert u_2(t)=-L\eta$.
\end{enumerate}

The sum of these actions produces in a period the displacement 
\begin{equation}
\Delta^- u=L\left(\eta \T -\frac{4F_--2F_+}{k}\right)
\label{eq:deltau}
\end{equation}
We notice that the strategy we just presented could be slightly improved by suitably modifying $\epsilon(t)$, for instance in a way to avoid the temporary forward movement of two of the legs. However these changes require an a priori knowledge of all the parameters of the systems, so that the strategy is, in a certain sense, calibrated to the situation, for instance requiring changes in $\dert \epsilon(t)$ exactly at the moment when the tension reaches the slip surface, i.e.~the boundary of $-C^*_B$. The strategy we presented instead shows the same behaviour for every choice of the parameters, provided that the assumption of large distortions is satisfied. 
Moreover we remark that such improvements of the strategy decrease only the numerator of the negative term inside the brackets in \eqref{eq:deltau}, so the main term is untouched and any improvement becomes negligible for large distortions $\eta \T$ or large stiffness $k$.

\begin{figure}[tb]
\centering
\subfloat[][]
{\begin{tikzpicture}[line cap=round,line join=round,>=stealth,x=1.0cm,y=1.0cm,line width=1.pt,scale=0.8]
\clip(-1,-0.5) rectangle (6.5,6.5);
\draw [->] (0.,0.) -- (6.,0.) node[anchor=north] {$\epsilon_1$};
\draw [->] (0.,0.) -- (0.,6.) node[anchor=east] {$\epsilon_2$};\footnotesize
\draw [<-,color=red] (1.,5.) -- (1,3) node [anchor=east] {$\gamma_3^+$}; 
\draw [<-,color=red] (1.,1.) --  (3,1) node [anchor=north] {$\gamma_2^+$};
\draw [<-,color=red] (5.,1.) --  (3,3) node [anchor=south west] {$\gamma_1^+$};
\draw [<-,color=red](1,3)-- (1.,1.);
\draw [<-,color=red](3,1)-- (5.,1.);
\draw [<-,color=red](3,3)-- (1.,5.);
\end{tikzpicture}}\quad
\subfloat[][]
{\begin{tikzpicture}[line cap=round,line join=round,x=1.cm,y=1.0cm,>=stealth,decoration={
    markings,
    mark=at position 0.5 with {\arrow{>}}},scale=1.6]
\clip(-2.,-1.5) rectangle (2.2,2.2);
\draw[line width=1.pt,color=red,fill=red,fill opacity=0.0] (1.,1.333) -- (0.3333,1.333) -- (-1.333,-0.3333) -- (-1.333,-1.) -- (0.3333,-1.) -- (1.,-0.3333) -- cycle;
\draw [<-] (2.,0.) -- (-2.,0.);
\draw [<-] (0.,2.) -- (0.,-1.5);\normalsize
\draw (2.,0.) node[anchor=north ] {$T_1$};
\draw (0.,2.) node[anchor=west] {$T_2$};
\draw[->,line width=1.2pt, dashed,postaction={decorate}] (0.3333,-1.) --(-0.833,0.166);
\draw[->,line width=1.2pt, dashed,postaction={decorate}] (-0.833,0.166) -- (1,0.166);
\draw[->,line width=1.2pt, dashed] (1,0.166) -- (1.,-0.3333);
\draw[->,line width=1.2pt, dashed,postaction={decorate}] (1.,-0.3333) -- (0.3333,-1.);
\draw  (1.,-0.3333) node[anchor=north] {$\gamma_3^+$};
\draw  (0.166,0.166) node[anchor=south ] {$\gamma_2^+$};
\draw (-0.666,-1) node[anchor=south west] {$\gamma_1^+$};
\end{tikzpicture}} 
\caption{Active distortion strategy \eqref{eq:example2} and associated evolution of the tension.}
\label{fig:ex2}
\end{figure}
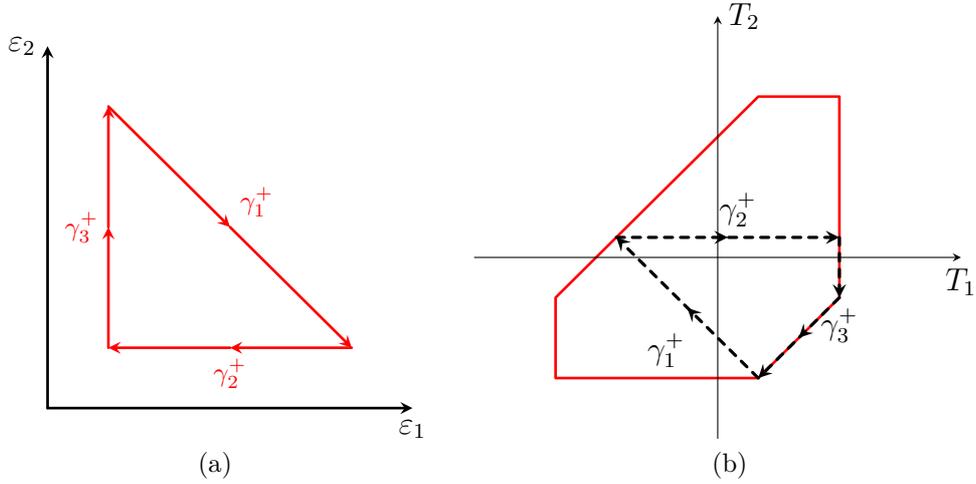

The history of active distortion \eqref{eq:example} was also chosen to show a backward movement of the crawler, that corresponds to proceeding in the direction of higher friction. A simple strategy to move forwards is given by the time reverse of strategy \eqref{eq:example}, namely
\begin{align}
\dert \epsilon_1(t)=\begin{cases}
\eta &\text{if $0<t<\tau$}\\
-\eta &\text{if $\tau<t<2\tau$}\\
0 &\text{if $2\tau<t<3\tau$}\\
\end{cases} &&
\dert \epsilon_2(t)=\begin{cases}
-\eta &\text{if $0<t<\tau$}\\
0 &\text{if $\tau<t<2\tau$}\\
\eta &\text{if $2\tau<t<3\tau$}\\
\end{cases}
\label{eq:example2}
\end{align}
Also in this case, after a preliminary stage, the tension configuration at the beginning of each period stabilizes to $T=-\beta_2$, that will be the starting condition in our analysis. The evolution of the tension is shown in Figure \ref{fig:ex2}.
After a period the displacement produced is
\begin{equation}
\Delta^+ u=L\left(\eta \T -\frac{5F_+-F_-}{2k}\right)
\label{eq:deltaup}
\end{equation}

We have that 
\begin{equation}
\Delta^+ u - \Delta^- u = \frac{9}{2} L (F_--F_+)>0
\end{equation}
and so  there is an advantage when moving in the direction of lower friction. This advantage becomes null as the ratio $F_+/F_-$ tends to one, while it increases to a constant when we approximate the threshold case $F_+/F_-=2$. 

We notice that the difference $\Delta^+ u - \Delta^- u$ between the displacement produced by our twin strategies does not depend on the amplitude $\eta \T$ of the distortion. This means that, if the crawler can produce only small distortions,  but slightly greater than the lower threshold  $(F_++F_-)/k$,  then a very large number of iterations of the first strategy is necessary to obtain a negative displacement equal to the positive one produced by a cycle of the second strategy. On the other hand, if the crawler can produce very large distortions (i.e.\ $\eta \T\to \infty$) the outcomes of the two strategies become comparable, in the sense that the ratio $\Delta^+ u / \Delta^- u$ tends to one.

We remark that reversing the strategy does not always reverse also the direction of motion, as it happens in the example above. A counterexample is given by the simple strategy
\begin{align}
\dert \epsilon_1(t)=\begin{cases}
\eta &\text{if $0<t<\tau$}\\
0 &\text{if $\tau<t<2\tau$}\\
-\eta &\text{if $2\tau<t<3\tau$}\\
\end{cases} &&
\dert \epsilon_2(t)=\begin{cases}
0&\text{if $0<t<\tau$}\\
\eta &\text{if $\tau<t<2\tau$}\\
-\eta &\text{if $2\tau<t<3\tau$}\\
\end{cases}
\label{eq:example3}
\end{align}
and its time-reverse, for sufficiently large distortions, namely $\eta \T >3F_-k$.
Both stategy \eqref{eq:example3} and its reverse produce the same, positive displacement after a period, equal to
\begin{equation}
\Delta u=L\frac{2F_--F_+}{k}
\end{equation}
We notice that in this case the displacement is independent of the distortion $\eta\T$, while with the previous strategies we had an asympotically linear growth in terms of $\eta\T$. The inefficiency of this strategies with respect to \eqref{eq:example2} can be seen intuitively also by looking at the behaviour of the crawler during a cycle. The first and the third legs perform both a forward and a backward movement, of amplitude growing with $\eta\T$, that almost cancel each other out, leaving only the final displacement $\Delta u$.

\medbreak

We conclude by remarking that the approach adopted in this paper can be extended also to analogous crawlers composed by a larger number of segments. 
Increasing the number of legs
enlarges the range of friction ratios under which motility in both directions is possible from 
$F_+ < F_- < 2 F_+$ to $F_+ < F_- < N F_+$.
Intuitively, a $N$-segment crawler can move each leg backwards  one by one, by leaning against the other $N-1$ legs, resulting in a strategy that generalizes \eqref{eq:example}. 
However the number of different scenarios that appear by varying the friction ratio
also increases with the number of segments, and a complete and detailed description of a generic evolution problem becomes soon burdensome.

\section*{Acknowledgement}
This work has been supported by the ERC Advanced Grant 340685-MicroMotility.



\end{document}